\documentclass[preprint,3p,times,onecolumn,sort&compress]{elsarticle}

\usepackage{color}
\usepackage{amssymb}
\usepackage{epstopdf}

\journal{Physica A}

	\usepackage{fancyheadings}
\renewcommand{\thispagestyle}[1]{} 

\begin{document}

\begin{frontmatter}



\title{Critical Temperature Studies of the Anisotropic Bi- and Multilayer Heisenberg Ferromagnets in Pair Approximation}

\author[label1]{Karol Sza\l{}owski\corref{label0}}
\ead{kszalowski@uni.lodz.pl}
\author[label1]{Tadeusz Balcerzak}
\address[label1]{Department of Solid State Physics, University of
{\L}\'{o}d\'{z}, ul. Pomorska 149/153, 90-236 {\L}\'{o}d\'{z},
Poland}
\cortext[label0]{corresponding author}

\begin{abstract}
The Pair Approximation method is applied to studies of the bilayer and multilayer magnetic systems with simple cubic structure. The method allows to take into account quantum effects related with non-Ising couplings. The paper adopts the anisotropic Heisenberg model for spin $S=1/2$ and considers the phase transition temperatures as a function of the exchange integrals strength in line with the role of intra- and interplanar anisotropic interactions in the onset of low-dimensional magnetism. The compensation effect for the Curie temperature is found for asymmetric interactions within the neighbouring planes of the bilayer system. The paper predicts the saturation of the Curie temperature for strong interplanar interactions. However, such an effect for the multilayer system occurs only when the interplanar interactions are of purely isotropic character. 
\end{abstract}


\end{frontmatter}



\section{Introduction}

Layered magnets, which are composed of some number of two-dimensional planes, have attracted considerable attention both theoretical and experimental \cite{deJongh1}. These systems bridge the gap between the two- and three-dimensional magnets, so that studying their properties provides some insight into the cross-over between the behaviours characteristic of various dimensionalities \cite{Araujo}. That makes the systems particularly interesting from the theoretical point of view. Furthermore, experimental realizations of the layered magnets may be based, for example, on organic compounds, which offer tunable magnetic properties and a huge variety of structures \cite{deJongh1,Sieklucka1,Balanda1,Jongh}.

The up to date available exact solutions for magnetic models are  limited to one-dimensional and some two-dimensional magnetic systems, which are mainly of the Ising type \cite{Baxter1}. The use of some general transformations allows to extend the range of soluble models \cite{Strecka2}. However, in order to study the transition between the two- and three-dimensional systems, the approximate methods are needed. The desired methods should be able to cope with the quantum effects resulting from the non-Ising couplings between the spins. Moreover, such methods should be in accordance with the Mermin-Wagner theorem \cite{MerminWagner1}, which imposes rather strict limitations on the magnetic ordering of low-dimensional magnets.

Numerous studies have  been mainly concentrated on bilayer ferromagnets, which are the first stage between the two-dimensional magnetic plane and three-dimensional bulk magnet. Among such systems, the bilayer Ising model on the square lattice has attracted considerable interest \cite{Horiguchi1,Ghaemi1,Li1,Hansen1,Lipowski1,Monroe1,Mirza1,Puszkarski1}. The studies included also other underlying planar lattices\cite{Sloutskin1}, especially the Bethe lattice \cite{Tucker1,Chin-Kin1,Albayrak1,Canko1}. Some works covered the interesting case of the spins larger that 1/2 \cite{Balcerzak4,Horiguchi1,Kapor}, as well as the spins different in both magnetic planes \cite{Balcerzak3,Jascur1,Albayrak2}. The considerations have been extended to include dilution \cite{Jascur3,Ainane1}, amorphous structure \cite{Bengrine1,Timonin}, and the case of unequal coupling strengths in each magnetic plane \cite{Oitmaa1}. The layered systems with anisotropic Heisenberg interactions have been studied in Refs.\cite{Jongh, Wei}. The properties of some more complicated multilayer structures were also investigated \cite{Lines1,Jascur2,Strecka1,Sy1,Lipowski2}. In addition, layered quantum Heisenberg antiferromagnets attract the attention, especially as they exhibit a phenomenon of quantum criticallity. For example, they are studied by means of Quantum Monte Carlo \cite{Sandvik1, Sandvik2} or renormalization-group methods \cite{Araujo}. However, according to our knowledge, there are only very few works able to describe (and compare) simultaneously within one method the single plane, bilayer, multilayer and  bulk systems.

In order to fill the gap our work describes the above systems in a thermodynamically consistent way. However, in this paper we have concentrated mainly on the ferromagnetic bilayer and multilayer systems consisting of two magnetically distinct kinds of magnetic planes. Adopting the anisotropic quantum Heisenberg model with spin 1/2, the description is based on the Pair Approximation \cite{Balcerzak2,Balcerzak1} method, which has been substantially modified in order to account for the anisotropic structure of the bilayer/multilayer system. We focus our study on the critical temperature, and emphasize the influence of various coupling strengths and the importance of interaction anisotropies. Let us note that the developed approach allows for a complete construction of the termodynamic description based on the expression for Gibbs free energy for the system.

The paper is organized as follows: in the theoretical section (II) we develop the Pair Approximation method for the anisotropic layered systems. In the third section (III) we present the results of numerical calculations for the phase transition temperature of the bilayer and multilayer systems. A comparison between these results as well as some discussion is offered there. In the last section (IV) the final remarks and conclusion are presented.

\section{Theoretical model}

A schematic picture of the bilayer and multilayer system in question is presented in the Fig.1. The bilayer system consists of two parallel planes, $A$ and $B$, containing  $A$ and $B$-type of atoms, respectively. We assume that the lattice sites form the simple cubic structure. Both intra- and interplanar exchange interactions are of the anisotropic Heisenberg type. We consider the case when $0\leq J^{\nu \mu}_{x}=J^{\nu \mu}_{y}=J^{\nu \mu}_{\perp} \le J^{\nu \mu}_{z}$ ($\nu=A,B; \, \mu =A,B$), which covers all intermediate situations between the pure Ising ($J^{\nu \mu}_{\perp}=0$) and the isotropic Heisenberg ($J^{\nu \mu}_{\perp}=J^{\nu \mu}_{z}$) models.

The multilayer system is built from the interacting bilayer segments by the infinite repetition of the bilayer in the direction perpendicular to the $A$ and $B$ planes.

\begin{figure}
\includegraphics[scale=0.26]{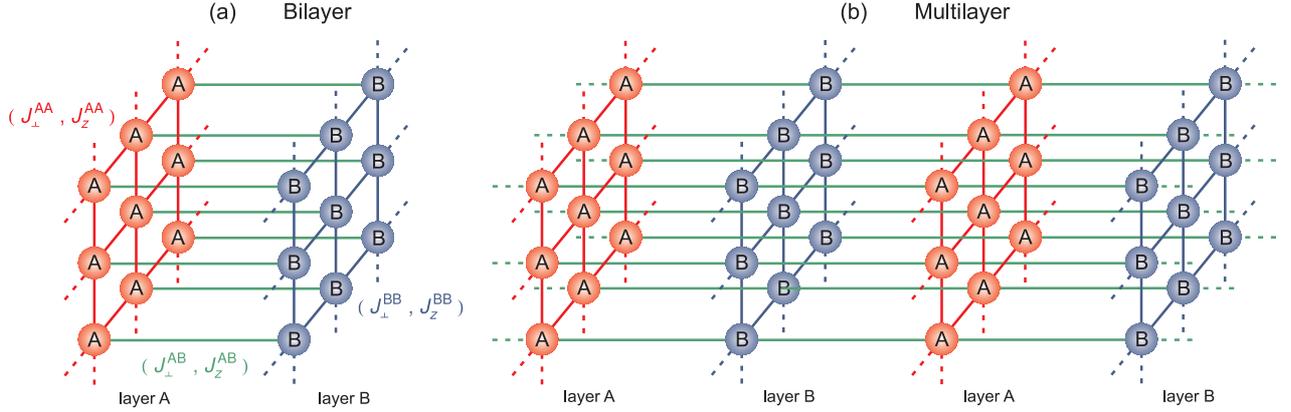}
\caption{(a) A schematic view of the bilayer composed of two layers, A and B. The intraplanar couplings are $J^{AA}_{x}=J^{AA}_{y}=J^{AA}_{\perp}$, $J^{AA}_{z}$ and $J^{BB}_{x}=J^{BB}_{y}=J^{BB}_{\perp}$, $J^{BB}_{z}$, respectively. The interplanar coupling is $J^{AB}_{x}=J^{AB}_{y}=J^{AB}_{\perp}$, $J^{AB}_{z}$. (b) A schematic view of a multilayer, containing an infinite number of subsequent layers A and B.}
\label{fig:fig1}
\end{figure}

The Hamiltonian of the bilayer can be written in the form of:
\begin{eqnarray}
\mathcal{H}&=&-\sum_{\left\langle i\in A,j\in B \right\rangle}^{}{\left[J_{\perp}^{AA}\left(S_{x}^{i}S_{x}^{j}+S_{y}^{i}S_{y}^{j}\right)+J_{z}^{AA}S_{z}^{i}S_{z}^{j}\right]}-\sum_{\left\langle i\in B,j\in B \right\rangle}^{}{\left[J_{\perp}^{BB}\left(S_{x}^{i}S_{x}^{j}+S_{y}^{i}S_{y}^{j}\right)+J_{z}^{BB}S_{z}^{i}S_{z}^{j}\right]}\nonumber\\
&&-\sum_{\left\langle i\in A,j\in B \right\rangle}^{}{\left[J_{\perp}^{AB}\left(S_{x}^{i}S_{x}^{j}+S_{y}^{i}S_{y}^{j}\right)+J_{z}^{AB}S_{z}^{i}S_{z}^{j}\right]}-h\sum_{i\in A}^{}{S^{i}_{z}}-h\sum_{i\in B}^{}{S^{i}_{z}}.
\label{eq1}
\end{eqnarray}
where $h$ stands for the external magnetic field. In the case of multilayer system the bilayer Hamiltonian should be summed up over all $A$ and $B$ layers, where each layer interacts with its neighbours from both sides.

The general method of constructing the thermodynamic description within PA has been sketched out in our previous papers \cite{Balcerzak2,Balcerzak1}. The crucial point of the PA is the method of construction of the single-site and pair density matrices, $\hat{\rho}^{i\in \nu}$ and $\hat{\rho}^{i\in \nu,j\in  \mu}$ ($\nu=A,B; \, \mu =A,B$), respectively. In the present case the matrices take the form of:
\begin{equation}
\hat{\rho}^{i\in \nu}=e^{\beta G^{\nu}}\exp\left[\beta\left(\Lambda^{\nu}+h\right)S^{i\in \nu}_{z} \right]
\label{rho1}
\end{equation}
($\nu=A,B$), and
\begin{equation}
\hat{\rho}^{i\in \nu,j\in  \mu}=e^{\beta G^{\nu \mu}}\exp \left(\, \beta\,  \mathcal{H}^{i\in \nu,j\in  \mu} \right) 
\label{rho2}
\end{equation}
where the pair Hamiltonian is of the form:
\begin{eqnarray}
\mathcal{H}^{i\in \nu,j\in  \mu}&=&J_{\perp}^{\nu \mu}\left(S_{x}^{i\in \nu}S_{x}^{j\in \mu}+S_{y}^{i\in \nu}S_{y}^{j\in \mu}\right)+J_{z}^{\nu \mu}S_{z}^{i\in \nu}S_{z}^{j\in \mu}\nonumber\\
&&+\left(\Lambda^{\nu \mu}+\Delta^{\nu \mu}/2+h\right)S^{i\in \nu}_{z}+\left(\Lambda^{\nu \mu}-\Delta^{\nu \mu}/2+h\right)S^{j\in \mu}_{z}
\label{Hnm}
\end{eqnarray}
($\nu=A,B; \, \mu=A,B$).

The single-site Gibbs energies, $G^{A}$ and $G^{B}$, which are obtained from the normalization condition for the single-site density matrices $\hat{\rho}^{i\in A}$ and $\hat{\rho}^{i\in B}$, respectively, are given by the expressions:
\begin{eqnarray}
G^{A}&=&-k_{\rm B}T \,\ln\left\{2\cosh\left[\frac{\beta}{2}\left(\Lambda^{A}+h\right)\right]\right\} 
\label{eq4a}
\\
G^{B}&=&-k_{\rm B}T \,\ln\left\{2\cosh\left[\frac{\beta}{2}\left(\Lambda^{B}+h\right)\right]\right\}. 
\label{eq4b}
\end{eqnarray}
On the other hand, the pair Gibbs energies, $G^{AA}, \, G^{BB}$ and $G^{AB}$, which are obtained from the normalization condition after diagonalization of the pair density matrices $\hat{\rho}^{i\in A,j\in A}, \, \hat{\rho}^{i\in B,j\in B}$ and $\hat{\rho}^{i\in A,j\in B}$, respectively, can be presented as:
\begin{eqnarray}
G^{AA}&=&-k_{\rm B}T \,\ln \left\{2\exp\left(\frac{\beta J^{AA}_{z}}{4}\right)\cosh\left[\beta\left(\Lambda^{AA}+h\right)\right]+2\exp\left(-\frac{\beta J^{AA}_{z}}{4}\right)\cosh\left(\frac{\beta}{2}J_{\perp}^{AA}\right)\right\} 
\label{eq5a}
\\
G^{BB}&=&-k_{\rm B}T \,\ln\left\{2\exp\left(\frac{\beta J^{BB}_{z}}{4}\right)\cosh\left[\beta\left(\Lambda^{BB}+h\right)\right]+2\exp\left(-\frac{\beta J^{BB}_{z}}{4}\right)\cosh\left(\frac{\beta}{2}J_{\perp}^{BB}\right)\right\} 
\label{eq5b}
\\
G^{AB}&=&-k_{\rm B}T \,\ln\left\{2\exp\left(\frac{\beta J^{AB}_{z}}{4}\right)\cosh\left[\beta\left(\Lambda^{AB}+h\right)\right]+2\exp\left(-\frac{\beta J^{AB}_{z}}{4}\right)\cosh\left[\frac{\beta}{2}\sqrt{\left(\Delta^{AB}\right)^2+\left(J_{\perp}^{AB}\right)^2}\,\right]\right\} 
\label{eq5c}.
\end{eqnarray}

Application of the PA method leads to the expression for the total Gibbs energy, from which all thermodynamic properties can self-consistently be derived. The general formula for the Gibbs energy has the form of:
\begin{equation}
G=\frac{1}{N}\sum_{\nu, \mu} \sum_{\left\langle i,j \right\rangle}\left[ G^{i\in \nu,j\in  \mu}-\frac{z-1}{z}\left(G^{i\in \nu}+G^{j\in \mu}\right)\right]
\label{gibbs}
\end{equation}
($\nu=A,B; \, \mu =A,B$), where $G^{i\in \nu}\equiv G^{\nu}$ are the single-site and $G^{i\in \nu,j\in  \mu}\equiv G^{\nu,\mu}$ are the pair Gibbs energies, respectively. The number of nearest neighbours (NN) is denoted by $z$ ($z=5$ for the bilayer and $z=6$ for the multilayer structure).

Taking into account the total number of NN pairs, the Gibbs energy per 1 lattice site for the bilayer system is given by:
\begin{equation}
G=G^{AA}+G^{BB}+G^{AB}/2-2\left(G^{A}+G^{B}\right)
\label{eq2}
\end{equation}
whereas for the multilayer system we obtain:
\begin{equation}
G=G^{AA}+G^{BB}+G^{AB}-\frac{5}{2}\left(G^{A}+G^{B}\right).
\label{eq3}
\end{equation}

The density matrices, and thus the Gibbs energy, depend on the parameters $\Lambda_{\nu}$, $\Lambda_{\nu,\mu}$ and $\Delta_{\nu,\mu}$, which can be expressed by means of four other parameters $\lambda^{\nu,\mu}$. The determination of these parameters is presented in full detail in the Appendix A. In principle, they are calculated from the variational minimization of the total Gibbs energy, which yields the conditions $\mathrm{Tr}\,\left(\hat{\rho}^{i\in \nu} S^{i\in \nu}_z\right)=\mathrm{Tr}\,\left(\hat{\rho}^{i\in \nu,j\in  \mu} S^{i\in \nu}_z\right)$. Having obtained the parameters, the Gibbs energies (\ref{eq2}) and (\ref{eq3}) and hence the total Gibbs energy (\ref{gibbs}) are at our disposal as the functions of arbitrary temperature ($\beta=1/k_{\rm B}T$) and the external magnetic field $h$. Let us emphasize that the local anisotropic exchange interactions ($J_{\perp}^{\nu \mu}$ and $J_{z}^{\nu \mu}$) are taken into account in the Eqs.(\ref{eq9a}) - (\ref{eq9d}) determining the parameters $\lambda^{\nu,\mu}$. Thus, these local equations give a possibility to study a variety of interesting cases for the intra- and interplanar couplings.

The main focus of the present work is the critical temperature $T_c$ of the bilayer and multilayer systems. The procedure used to determine $T_c$ (in case of the second-order phase transition) for the systems of interest is presented in the Appendix B. In a general case, calculation of $T_c$ requires solving a determinant equation (\ref{eq10}) of the size $4\times 4$. However, in the particular case when the two layers $A$ and $B$ are magnetically equivalent, i.e., for $J^{AA}_{z}=J^{BB}_{z}=J_{z}$, \, $J^{AA}_{\perp}=J^{BB}_{\perp}=J_{\perp}$, we obtain the simple formulas for the critical temperature:
\begin{equation}
4C+C^{AB}=3
\label{eq13}
\end{equation}
for the bilayer system, and 
\begin{equation}
2C+C^{AB}=2
\label{eq14}
\end{equation}
for the multilayer system.
The $C$ and $C^{AB}$ coefficients take the form of:
\begin{eqnarray}
C&=&\exp\left(-\frac{J_{z}}{2k_{\rm B}T_c}\right)\cosh\left(\frac{J_{\perp}}{2k_{\rm B}T_c}\right)\nonumber\\
C^{AB}&=&\exp\left(-\frac{J_{z}^{AB}}{2k_{\rm B}T_c}\right)\cosh\left(\frac{J_{\perp}^{AB}}{2k_{\rm B}T_c}\right)
\label{eq15}
\end{eqnarray}
On the basis of Eqs.(\ref{eq13}) and (\ref{eq14}) it is a simple task to check that, for instance, for the uncoupled Ising layers 
(when $J_{\perp}=0$ and $J^{AB}_{\perp}=J^{AB}_{z}=0$) we obtain: $k_{\rm B}T_{c}/J_{z}=1/2\ln2$. On the other hand, when the Ising planes are coupled, the result is $k_{\rm B}T_{c}/J_{z}=1/2\ln[z/(z-2)]$ where $z=5$ is the NN number for the bilayer system, and $z=6$ stands for the multilayer system. We see that the result for the uncoupled planes corresponds to the Curie temperature of one monolayer with s.q. structure ($z=4$). These analytical solutions are in agreement with the findings of previous papers on the PA method \cite{Balcerzak6, Balcerzak2}. 

In other limiting case, when the system contains only isotropic Heisenberg interactions ($J_{\perp}=J_{z}=J$), and assembles uncoupled planes, we obtain $k_{\rm B}T_{c}/J_{z}=0$ which is in agreement with the Mermin-Wagner theorem \cite{MerminWagner1}. On the other hand, for the interacting Heisenberg planes we obtain $k_{\rm B}T_{c}/J_{z}=1/\ln[z/(z-4)]$, where $z=5$ stands for the bilayer and $z=6$ corresponds to the multilayer system. Again, this result is in agreement with previous studies of the Heisenberg systems within the PA method \cite{Balcerzak6}.

From the above remarks one can conclude that the PA method should be able to reveal the onset of magnetism in the case of the isotropic Heisenberg planes ($J^{AA}_{z}=J^{BB}_{z}=J^{AA}_{\perp}=J^{BB}_{\perp}=J$) when the interplanar interactions are gradually being switched on. It turns out that either from Eq.(\ref{eq13}) or (\ref{eq14}), when $J^{AB}_{z} \to 0$, we find the common result:
\begin{equation}
\frac{k_{\rm B}T_{c}}{J}=-\frac{1}{\ln\left(J^{AB}_{z}/J\right)}
\label{eq16}
\end{equation}
It is interesting to see that in the logarithmic law (\ref{eq16}) the perpendicular interaction between the planes
($J^{AB}_{\perp}$) has no influence on the Curie temperature.

Finally, for the non-interacting planes ($J^{AB}_{\perp}=J^{AB}_{z}=0$), when we define the intraplanar anisotropy parameter as: $\delta=(J_{z}-J_{\perp})/J_{z}$, we find that:
\begin{equation}
\frac{k_{\rm B}T_{c}}{J_{z}}=\frac{1}{\ln\left(1/\delta\right)}
\label{eq17}
\end{equation}
for $\delta \to 0$. This result is in agreement with our previous findings for the anisotropic Heisenberg model in the diluted systems \cite{Balcerzak1}, when the effective coordination number amounts $z_{\rm eff}=4$ . It describes the onset of 2D magnetism as a result of anisotropic interactions.

The numerical results based on the general equation for the critical temperature (\ref{eq10}), especially where the analytical solutions do not exist, are presented in the next section.

\section{Numerical results and discussion}

\begin{figure}
\includegraphics[scale=0.70]{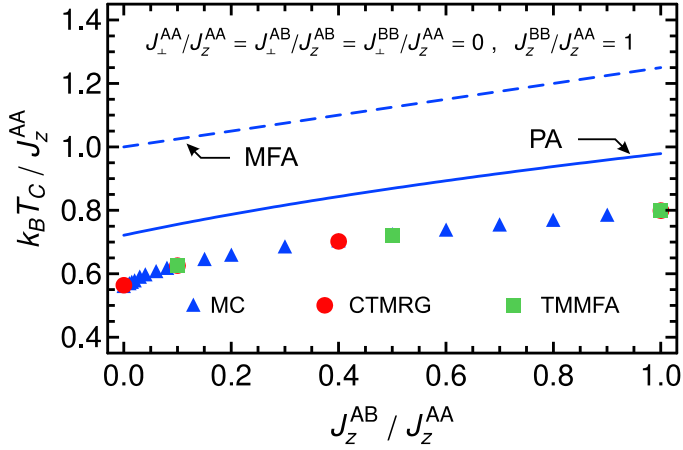}
\caption{The critical temperature of a bilayer system with Ising interlayer and intralayer couplings as a function of relative strength of interlayer interaction. Dashed line is the MFA result, solid line - PA result. Triangles denote the results of Monte Carlo method (Ref.~\cite{Hansen1}), circles - Corner Transfer Matrix Renormalization Group (Ref.~\cite{Li1}), squares - Transfer Matrix Mean Field Approximation (Ref.~\cite{Lipowski1}).}
\label{fig:fig2}
\end{figure}

First let us study the case of the critical temperature of the bilayer system composed of two magnetic planes with interplanar couplings of the Ising type and equal exchange integral strength ($J^{BB}_{z}/J^{AA}_{z}=1$, $J^{BB}_{\perp}/J^{BB}_{z}=J^{AA}_{\perp}/J^{AA}_{z}=0$). The critical temperature is plotted in the Fig.~\ref{fig:fig2} vs. $J^{AB}_{z}/J^{AA}_{z}$. The linear dependence predicted by the Mean Field Approximation, $k_{\rm B}T^{MFA}_{c}=J^{AA}_{z}+J^{AB}_{z}/4$, is shown by the dashed line, while the solid line presents the result of the present (PA) method. The closed symbols depict the predictions of Monte Carlo simulations (after Ref.~\cite{Hansen1}), Transfer Matrix Mean Field Approximation (after Ref.~\cite{Lipowski1}) and Corner Transfer Matrix Renormalization Group (after Ref.~\cite{Li1}). It is visible that the MFA prediction significantly overestimates $T_c$, while the PA results, which form a nonlinear dependence, are much closer to the above mentioned accurate estimations of the critical temperature.

\begin{figure}
\includegraphics[scale=0.70]{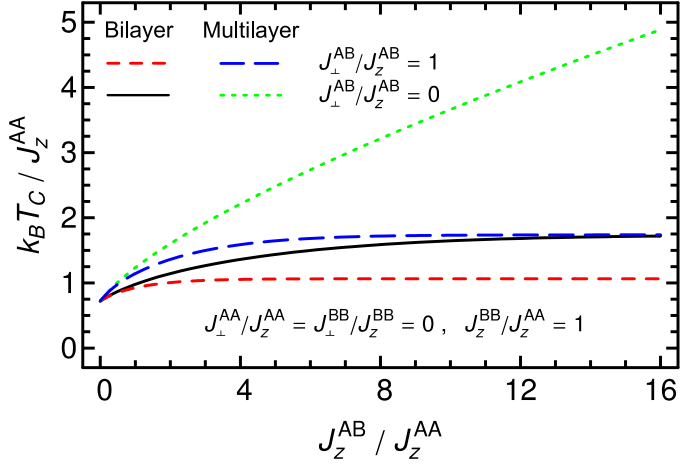}
\caption{The critical temperature of the bilayer and multilayer system with equal Ising couplings within the magnetic planes as a function of the relative strength of interlayer interaction. The interlayer interaction is either of the Ising or isotropic Heisenberg type.}
\label{fig:coupling1}
\end{figure}
\begin{figure}
\includegraphics[scale=0.70]{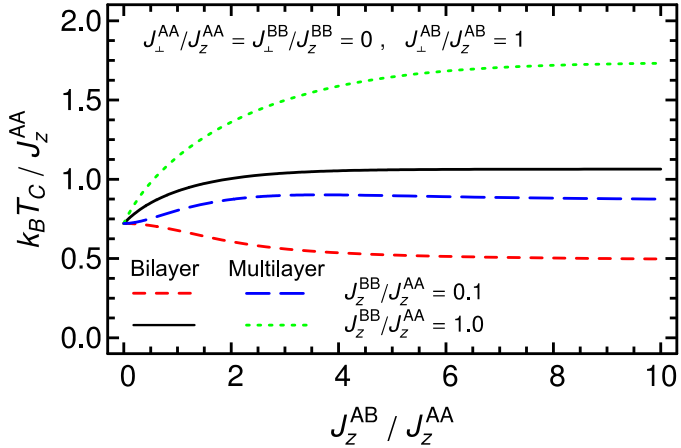}
\caption{The critical temperature of the bilayer and multilayer system with equal and unequal Ising couplings within the magnetic planes as a function of the relative strength of interlayer interaction. The interlayer interaction is of the isotropic Heisenberg type.}
\label{fig:coupling2}
\end{figure}

\begin{figure}
\includegraphics[scale=0.70]{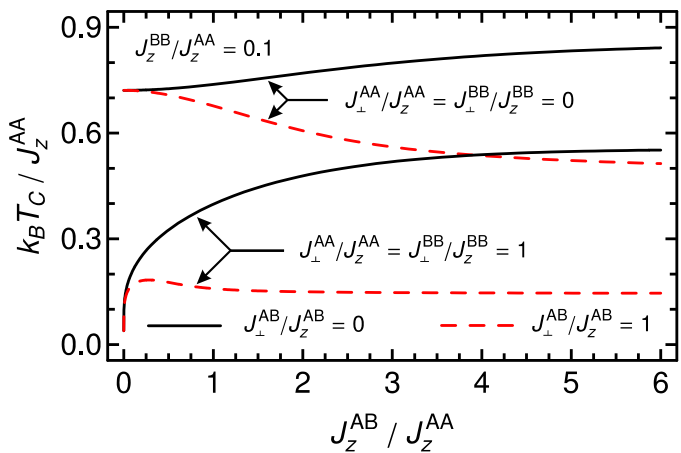}
\caption{The critical temperature for a bilayer system with unequal strength of intralayer couplings in both planes, either for the Ising or isotropic Heisenberg coupling between the planes.}
\label{fig:coupling3}
\end{figure}

An influence of the increasing interplanar coupling on the critical temperature of the bilayer or multilayer system is illustrated in the Fig.~\ref{fig:coupling1}. The plots  show the dependence of $k_{\rm B}T_{c}/J^{AA}_{z}$ on $J^{AB}_{z}/J^{AA}_{z}$. It is assumed that the intraplanar interactions within A and B planes are of purely Ising type ($J^{AA}_{\perp}=J^{BB}_{\perp}=0$) and they are taken with equal strength ($J^{AA}_{z}=J^{BB}_{z}$). The AB couplings are assumed to be either of Ising or isotropic Heisenberg type. All the curves start from a common point $k_{\rm B}T_{c}/J^{AA}_{z}=1/2 \ln 2$, which constitutes the critical temperature of a single Ising plane. Both for bi- and multilayer, the Ising coupling elevates the Curie temperature stronger that the isotropic Heisenberg coupling. For the case of a bilayer system, the critical temperature tends to saturate even though the coupling $J^{AB}_{z}/J^{AA}_{z}$ is increased to infinity. The saturation value for the Ising coupling is found as $k_{\rm B}T_{c}/J^{AA}_{z}=1/2 \ln\left(4/3\right)$, whereas the corresponding limiting value for strong isotropic Heisenberg coupling is $k_{\rm B}T_{c}/J^{AA}_{z}=1/2\ln\left(8/5\right) $. The tendency for the critical temperature to saturate is also clear for the case of the multilayer system with the isotropic Heisenberg AB interactions. Interestingly, the limiting critical temperature of a strongly Ising-coupled bilayer system is the same as for the strongly Heisenberg-coupled multilayer system. On the contrary, for the multilayer system with pure Ising coupling the Curie temperature tends to increase unlimitedly when $J^{AB}_{z}/J^{AA}_{z}$ is increased. Such behaviour of the curves is in agreement with Eqs.(\ref{eq13}) and (\ref{eq14}).

The same kind of behaviour has been detected for the case of a bilayer and multilayer system for which the intraplanar interactions are of isotropic Heisenberg type. For this reason such temperatures have not been illustrated in a separate plot. The only qualitative difference is that all the curves commence at $T_c$ equal to zero for uncoupled planes. The unlimited increase of the critical temperature takes place only for the multilayer system with Ising coupling. Also, the critical temperatures of the bilayer system with the Ising interlayer coupling and for the Heisenberg-coupled multilayer system tend to a common limit, as for the previously described case.

The influence of unequal strength of the Ising intraplanar coupling within the layers A and B on the behaviour of bilayer and multilayer system is presented in the Fig.~\ref{fig:coupling2}. 
In such asymmetric situation the Curie temperatures have to be obtained from the full determinant (\ref{eq10}).
It is visible that in all the cases considered in Fig.~\ref{fig:coupling2} the critical temperature tends to saturate for strong interlayer coupling. However, in the case of the bilayer with $J^{BB}_{z}/J^{AA}_{z}=0.1$ and strong coupling the limiting critical temperature  is significantly lower than $T_c$ for the single A plane with Ising interactions. This suggests that in such a case the presence of the Heisenberg lateral bounds  weakens the ferromagnetic order in the planar system. An analogous effect is absent in the multilayer system with $J^{BB}_{z}/J^{AA}_{z}=0.1$. Moreover, for equal strength of intraplanar couplings ($J^{BB}_{z}/J^{AA}_{z}=1$), the critical temperature of multilayer system can be elevated as a result of the presence of Heisenberg isotropic coupling between the magnetic planes.  

To emphasize the influence of the isotropic Heisenberg coupling on the critical temperature for a bilayer system with unequal intraplanar couplings, we present the next Fig.~\ref{fig:coupling3}, for $J^{BB}_{z}/J^{AA}_{z}=0.1$. It can be observed that in the case of the Ising coupling within the planes ($J^{AA}_{\perp}/J^{AA}_{z}=J^{BB}_{\perp}/J^{BB}_{z}=0$), the Ising interplanar coupling causes the critical temperature to increase, while the isotropic Heisenberg coupling reduces $T_c$. The critical temperature remains reduced even for a strong coupling. The situation is qualitatively different for the case of the bilayer system with isotropic Heisenberg interactions within the planes ($J^{AA}_{\perp}/J^{AA}_{z}=J^{BB}_{\perp}/J^{BB}_{z}=0$). For the uncoupled system, the critical temperature is equal to zero. Initially, introducing either the Ising or isotropic Heisenberg coupling $T_c$ becomes elevated. For the Ising AB interaction, the temperature keeps increasing monotonically and tends to saturate for strong coupling. On the contrary, for the Heisenberg interplanar coupling, $T_c$ reaches a certain maximum value and then becomes reduced until it reaches some limiting value (quite low when compared to the case of the Ising coupling).

\begin{figure}
\includegraphics[scale=0.70]{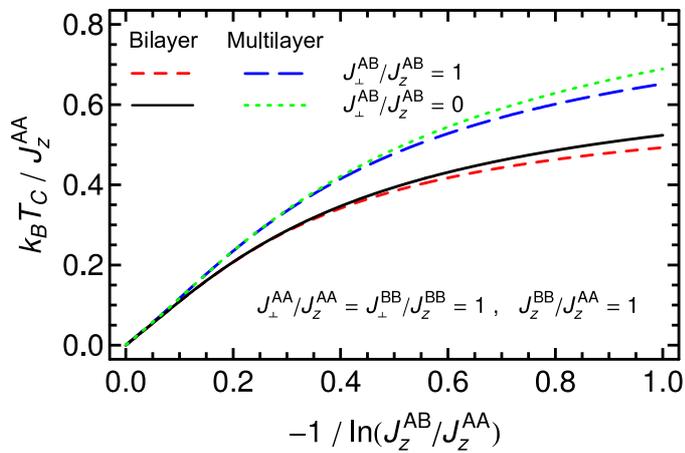}
\caption{The critical temperature for the bilayer and multilayer system with isotropic Heisenberg intralayer coupling as a function of inverse logarithm of the relative interplanar exchange interaction. The interplanar coupling is either of Ising or isotropic Heisenberg type.}
\label{fig:heisenberg1}
\end{figure}

The behaviour of the critical temperature of the bilayer and multilayer system with isotropic Heisenberg intraplanar couplings deserves particular attention. As stated in the theoretical part a single plane with isotropic Heisenberg interactions does not show magnetic ordering according to the Mermin-Wagner theorem \cite{MerminWagner1,Hohenberg1}. Therefore, the rise of ordered state as a consequence of introducing the interaction between the planes appears especially interesting. In the Fig.~\ref{fig:heisenberg1} the critical temperature $k_{\rm B}T_{c}/J^{AA}_{z}$ is shown as a function of the inverse logarithm of the couplings ratio $J^{AB}_{z}/J^{AA}_{z}$. Both the case of bilayer and multilayer system is taken into account. Moreover, the AB coupling is assumed either in the Ising or isotropic Heisenberg form. It is visible that for both systems and both kinds of interplanar interactions the critical temperature follows the same asymptotic behaviour when $J^{AB}_{z}/J^{AA}_{z}$ vanishes. It has been shown analytically in the previous section that this asymptotic behaviour takes the form of $k_{\rm B}T_{c}/J^{AA}_{z}=-1/\ln\left(J^{AB}_{z}/J^{AA}_{z}\right)$. For stronger coupling it is visible that the interplanar interaction has a more pronounced effect on the multilayer than on the bilayer system. Moreover, the Ising coupling causes the critical temperature to increase more rapidly than the isotropic Heisenberg one. Let us note that such kind of behaviour predicted here by the PA method is consistent with the expectations based on the scaling theory  \cite{Pokrovsky1} as well as with some other approaches \cite{Irkhin1,Katanin1,Araujo}. It bears some resemblance to the result for a single Heisenberg plane, in which the interaction anisotropy in spin space breaks the assumptions of the Mermin-Wagner theorem, and for which case the PA method predicts  $T_c\propto -1/\ln\left[\left(J_{z}-J_{\perp}\right)/J_{z}\right]$  (see Eq.(\ref{eq17}) and Ref.~\cite{Balcerzak1}). This result is in agreement with Ref.~\cite{Araujo}.

\begin{figure}
\includegraphics[scale=0.70]{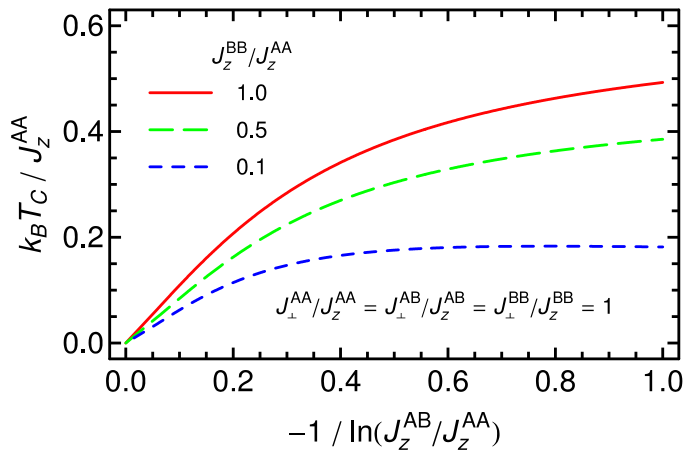}
\caption{The critical temperature for a bilayer system with isotropic Heisenberg intralayer coupling, as a function of inverse logarithm of the relative interplanar coupling energy. The interplanar coupling is of isotropic Heisenberg type. Either equal or unequal intralayer couplings are assumed.}
\label{fig:heisenberg2}
\end{figure}

The asymptotic behaviour of the critical temperature can also be tested for the case of unequal couplings in both layers A and B with isotropic Heisenberg-type interactions. Such a situation for the bilayer system is presented in the Fig.~\ref{fig:heisenberg2}. It follows that the proportionality of $T_c$ to $1/\ln\left(J^{AB}_{z}/J^{AA}_{z}\right)$ is unchanged by weakening the relative strength of coupling within the B layer; however, the coefficient of this proportionality decreases.

\begin{figure*}
\includegraphics[scale=0.69]{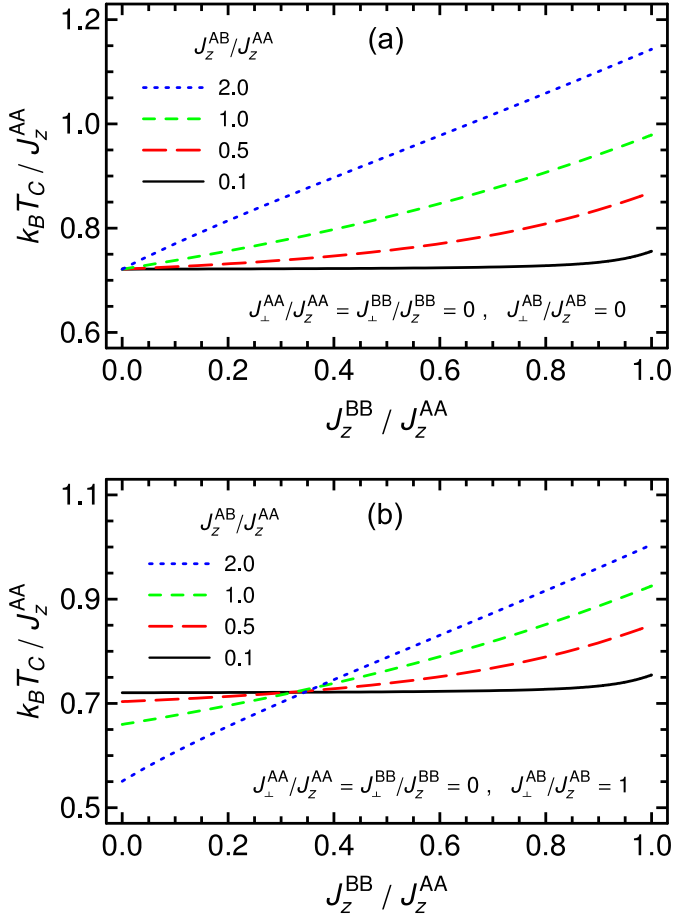}
\caption{The normalized critical temperature for a bilayer system as a function of the relative strength of the Ising couplings within A and B planes for various interplanar coupling energies: (a) Ising coupling between the planes; (b) isotropic Heisenberg coupling between the planes.}
\label{fig:compensation1}
\end{figure*}

The case of unequal strength of the intraplanar AA and BB couplings may have also other important consequences. For instance, the behaviour of the critical temperature as a function of the ratio $J^{BB}_{z}/J^{AA}_{z}$ for various interplanar AB couplings is especially worth mentioning. Let us fix the value of $J^{AA}_{z}$ and normalize the critical temperature to it. In the Fig.~\ref{fig:compensation1}(a) the dependence of the critical temperature on the ratio of intraplanar couplings $J^{BB}_{z}/J^{AA}_{z}$ is presented for selected values of the pure Ising-like interaction between the planes, i.e. when $J^{AB}_{\perp}=0$. It is visible that within the limit of vanishing intraplanar couplings $J_{z}^{BB}$ the critical temperature of the system tends to the value predicted for the sole A magnetic plane, i.e. $k_{\rm B}T_{c}/J^{AA}_{z}=1/2\ln 2$, independently on the energy of the interplanar interaction. The increase of $J^{BB}_{z}$ elevates $T_{c}$ and the effect is more rapid for stronger coupling between the planes. 

However, the situation noticeably changes when the pure Ising interplanar coupling is replaced by the isotropic Heisenberg one (i.e., for $J^{AB}_{\perp}=J^{AB}_{z}$), which is depicted in the Fig.~\ref{fig:compensation1} (b).  For a very weak or even vanishing coupling in the plane B, the effect of Heisenberg coupling $J^{AB}_{\perp}$ on the critical temperature is detrimental; a stronger $J^{AB}_{\perp}$ coupling reduces significantly the critical temperature below the value of $k_{\rm B}T_{c}/J^{AA}_{z}=1/2 \ln 2$ which is characteristic of a single A plane. Such a decrease can be interpreted as an effect of the coupling between the perpendicular components of the spins, tending to destroy the spin order along the $z$ axis, and is indeed a quantum effect. This demonstrates that, in certain circumstances, the ferromagnetic coupling is able to reduce the value of the critical temperature. As it is visible in Fig.~\ref{fig:compensation1} (b) this reduction of $T_{c}$ becomes less pronounced for the increasing coupling in the B plane. On the other hand, when $J^{BB}_{z}$ is comparable with $J^{AA}_{z}$  the behaviour of the critical temperature follows the trend presented in the Fig.~\ref{fig:compensation1}(a). It is particularly interesting that at some value of $J^{BB}_{z}/J^{AA}_{z}$ the critical temperature $k_{\rm B}T_{c}/J^{AA}_{z}=1/2\ln 2$ is restored for each curve, even though the disordering Heisenberg interplanar couplings are present. The value of $J^{BB}_{z}/J^{AA}_{z}$ at which the compensation of the opposite influences of $J^{AB}_{\perp}$ and $J^{BB}_{z}$ takes place is very weakly sensitive to the strength of interplanar coupling. This is visible in the Fig.~\ref{fig:compensation1} (b) as an approximate interception point of all the curves plotted for various values of $J_{z}^{AB}/J^{AA}_{z}$.

\begin{figure}
\includegraphics[scale=0.70]{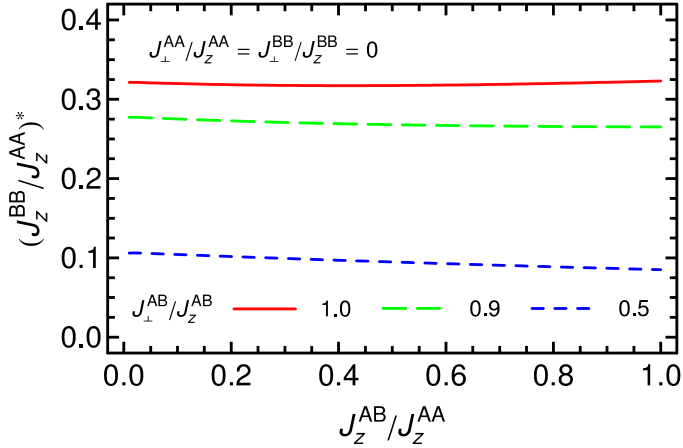}
\caption{The position of the compensation point for the bilayer system with the Ising coupling within the planes as a function of the relative strength of isotropic Heisenberg interlayer interaction.}
\label{fig:compensation2}
\end{figure}

The insensitivity of the coupling ratio $\left(J^{BB}_{z}/J^{AA}_{z}\right)^{*}$, for which $T_c$ has the constant value (characteristic of an uncoupled bilayer system), on the interplanar interaction is further illustrated in the Fig.~\ref{fig:compensation2}. There, the values of $\left(J^{BB}_{z}/J^{AA}_{z}\right)^{*}$ are plotted as a function of the interplanar coupling $J^{AB}_{z}/J^{AA}_{z}$ strength. The plot is prepared not only for isotropic Heisenberg interplanar coupling ($J^{AB}_{\perp}/J^{AB}_{z}=1$), but also for two smaller values of interplanar interaction $J^{AB}_{\perp}$. It is noticeable that the curves are almost flat (which is most evident for the isotropic Heisenberg AB interaction) so that the compensation point $\left(J^{BB}_{z}/J^{AA}_{z}\right)^{*}$ is quite well defined for a wide range of $J^{AB}_{\perp}$. The plot also shows that the value of $\left(J^{BB}_{z}/J^{AA}_{z}\right)^{*}$ is shifted towards lower values when $J^{AB}_{\perp}/J^{AB}_{z}$ decreases. In the limiting case of Ising AB coupling, i.e., for $J^{AB}_{\perp}=0$, the 'compensation' takes place at $J^{BB}_{z}/J^{AA}_{z}=0$, which is just the result illustrated in the Fig.~\ref{fig:compensation1}(a).

\begin{figure}
\includegraphics[scale=0.70]{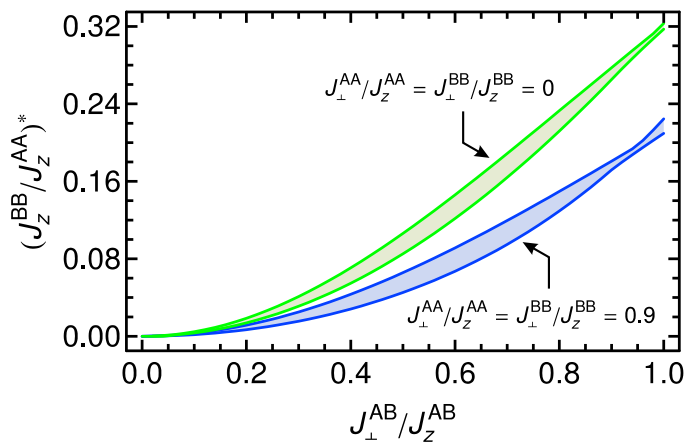}
\caption{The diffusion of the compensation point for a bilayer system as a function of the interlayer interaction anisotropy. The intraplanar couplings were assumed either as the Ising or anisotropic Heisenberg ones.}
\label{fig:compensation3}
\end{figure}

Finally, let us study how the relative amount of perpendicular component $J^{AB}_{\perp}$ in the AB couplings influences the positional diffusion of the compensation point. Fig.~\ref{fig:compensation3} presents the values of $\left(J^{BB}_{z}/J^{AA}_{z}\right)^{*}$ for $J^{AB}_{\perp}/J^{AB}_{z}$ varying from 1 (isotropic Heisenberg AB coupling) down to 0 (Ising coupling). For each particular value of $J^{AB}_{\perp}/J^{AB}_{z}$ all the results $\left(J^{BB}_{z}/J^{AA}_{z}\right)^{*}$, which are adequate for $J^{AB}_{z}/J^{AA}_{z}$ ranging from 0 to 1 (i.e. covering the range of the Fig.~\ref{fig:compensation2}), are shown by the shadowed area. The 'thickness' of this shadowed area plotted in the Fig.~\ref{fig:compensation3} corresponds then to the variation of $\left(J^{BB}_{z}/J^{AA}_{z}\right)^{*}$ as a function of $J^{AB}_{z}/J^{AA}_{z}$. Thus, the smallest thickness means the weakest sensitivity of the compensation point to the strength of interplanar coupling. It is visible that the compensation points are only slightly sensitive to AB coupling strength either for the isotropic Heisenberg interaction or when $J^{AB}_{\perp}\to 0$, while for the intermediate values of $J^{AB}_{\perp}/J^{AB}_{z}$ the compensation is considerably diffused. 

In the same picture (Fig.~\ref{fig:compensation3}) we also included the case when the intraplanar interactions within A and B planes are not purely of the Ising type. It is observable that a similar compensation takes also place when $J^{AA}_{\perp}/J^{AA}_{z}=J^{BB}_{\perp}/J^{BB}_{z}>0$, but the effect is less pronounced and the value of $\left(J^{BB}_{z}/J^{AA}_{z}\right)^{*}$ needed to compensate the influence of AB coupling on the critical temperature is lower. We have observed that the critical temperature of a bilayer ferromagnetic system undergoes the strongest reduction when the interplanar coupling is of the isotropic Heisenberg type and the intraplanar coupling within the planes is of the Ising type. Let us also mention that the compensation phenomenon is characteristic of a bilayer system with $J^{AB}_{\perp}/J^{AB}_{z}>0$, and no similar effect has been observed for a multilayer system, when the system  consists of an infinite number of subsequent A and B planes. 

\section{Final remarks and conclusion}

It has been found that the PA method is useful for the magnetic studies of both the bilayer and multilayer systems. In particular, the ferromagnetic phase transition temperature has been thoroughly examined. The anisotropic Heisenberg model has been adopted with the variable exchange integral between the planes.

The model allows to control two limiting cases of the system: the first case is when the planes are totally separated and their properties are purely 2-dimensional. The second limiting case comprises situation when the system becomes an infinite bulk crystal. The application of the PA method for those limits gives a good consensus with the results existing in the literature. In particular, for 2D Heisenberg systems the agreement with Mermin-Wagner theorem has been found.

For an intermediate situation when the bilayer or multilayer system is considered, the PA method allows to take into account not only the magnetic anisotropy in  spin space (with various $J_{\perp}$ and $J_{z}$), but also the directional anisotropy (different $J^{AA}$, $J^{BB}$ and $J^{AB}$) as well. In particular, it enables to study the layers with asymmetric exchange interactions ($J^{AA}\ne J^{BB}$). Thus, the current approach is formulated more generally than that presented in the previous work (Ref.~\cite{Balcerzak2}).

Regarding the main results obtained in the paper, the asymptotic law for the Curie temperature decay (Eq. \ref{eq16}), when the interplanar coupling decreases, has been derived. It has been found that only the Ising-type coupling influences the Curie temperature in such asymptotic law, and the formula is common both for the bilayer and multilayer system. This result of the PA method is in agreement with the predictions of
the scaling theory  \cite{Pokrovsky1} as well as with some other approaches \cite{Irkhin1,Irkhin2,Katanin1,Araujo}.

Another important result concerns the asymptotic behaviour of the Curie temperature when the interplanar coupling increases to infinity. It has been found that for the bilayer with intraplanar interactions both of the Ising and Heisenberg-type, when the interplanar (either Ising or Heisenberg) coupling  increases, the Curie temperature reaches its upper limit.  The limiting values can be found on the basis of Eq.(\ref{eq13}). On the other hand, for the multilayer system with  intraplanar interactions both of the Ising and Heisenberg-type, the saturation of the Curie temperature takes place only when the interplanar couplings are of the Heisenberg-type.  In this case, the limiting values of the Curie temperature can be found from Eq.(\ref{eq14}).

Finally, for the bilayer system with asymmetric intraplanar interactions ($J^{AA}>J^{BB}$) the 'compensation point' for the Curie temperature is found, as shown in the Fig.\ref{fig:compensation1} (b). This effect arises from some decreasing of the Curie temperature when the $J^{BB}$-interaction decreases, provided the interplanar couplings are of the Heisenberg type. Such an effect has not been found in the multilayer system. It reflects the fact that the properties of multilayer system cannot be considered as a simple multiplication of the bilayer properties, but they can be substantially  different. In the view of that finding a confirmation of the 'compensation point' existence by other methods, for instance, by the quantum MC simulations would be desirable.

Let us remark that according to the theoretical part the application of the PA method far exceeds the calculations of the phase transition point. Since the Gibbs energy is at our disposal, the calculations of all thermodynamic quantities for the system in question is possible. For instance, the studies of the thermodynamic response functions, such as the susceptibility or specific heat might be of interest. However, in our opinion, such studies can be done in a separate paper. 

The method can be developed to include antiferromagnetic interactions. Further extensions may also comprise the structurally disordered systems as well as the models containing higher spins.

\section*{Acknowledgments}
The numerical calculations and a part of analytical derivations have been performed with the Wolfram Mathematica 8.0.1 software \cite{Wolfram1}. 

This work has been supported by Polish Ministry of Science and Higher Education by a special purpose grant to fund the research and development activities and tasks associated with them, serving the development of young scientists and doctoral students.

\appendix
\section{Determination of variational parameters}

The molecular field parameters $\Lambda$ and $\Delta$ are specified separately for the bilayer and multilayer system, taking into account the different number of NN bounds. For instance, for the bilayer system:
\begin{eqnarray}
\Lambda^{A}&=&4\lambda^{AA}+\lambda^{AB}\nonumber\\
\Lambda^{B}&=&4\lambda^{BB}+\lambda^{BA}\nonumber\\
\Lambda^{AA}&=&3\lambda^{AA}+\lambda^{AB}\nonumber\\
\Lambda^{BB}&=&3\lambda^{BB}+\lambda^{BA}\nonumber\\
\Lambda^{AB}&=&2\left(\lambda^{AA}+\lambda^{BB}\right)\nonumber\\
\Delta^{AB}&=&4\left(\lambda^{AA}-\lambda^{BB}\right)\nonumber\\
\Delta^{AA}&=&\Delta^{BB}=0.
\label{eq6}
\end{eqnarray}
In turn, for the multilayer system the parameters are of the form:
\begin{eqnarray}
\Lambda^{A}&=&4\lambda^{AA}+2\lambda^{AB}\nonumber\\
\Lambda^{B}&=&4\lambda^{BB}+2\lambda^{BA}\nonumber\\
\Lambda^{AA}&=&3\lambda^{AA}+2\lambda^{AB}\nonumber\\
\Lambda^{BB}&=&3\lambda^{BB}+2\lambda^{BA}\nonumber\\
\Lambda^{AB}&=&2\left(\lambda^{AA}+\lambda^{BB}\right)+\frac{1}{2}\left(\lambda^{AB}+\lambda^{BA}\right) \nonumber\\
\Delta^{AB}&=&4\left(\lambda^{AA}-\lambda^{BB}\right)+\lambda^{AB}-
\lambda^{BA}\nonumber\\
\Delta^{AA}&=&\Delta^{BB}=0.
\label{eq7}
\end{eqnarray}
The four parameters $\lambda^{\nu \mu}$ ($\nu=A,B; \, \mu =A,B$) introduced in Eqs.(\ref{eq6}) and (\ref{eq7})  have a meaning of the one-bond molecular fields, i.e., the fields acting on the $\nu$-atom and originating from its $\mu$-neighbour. These parameters can be obtained from the variational principle for the free energy:
\begin{equation}
\frac{\partial G}{\partial \lambda^{\nu \mu}}=0
\label{eq8}
\end{equation}

Let us remark that in the PA method the variational principle \ref{eq8} is equivalent to the condition that the mean value of a given spin $\left< S^{i\in \nu}_z\right>$ (magnetization) remains the same when it is calculated either from $\hat{\rho}^{i\in \nu}$ or $\hat{\rho}^{i\in \nu,j\in  \mu}$ density matrix i.e., $\mathrm{Tr}\,\left(\hat{\rho}^{i\in \nu} S^{i\in \nu}_z\right)=\mathrm{Tr}\,\left(\hat{\rho}^{i\in \nu,j\in  \mu} S^{i\in \nu}_z\right)$. We emphasize that this condition is vital to the correct description of the system, since it guarantees the consistence between the quantum state of a single lattice site described either by single-site or by pair density matrix. Let us mention that for spin 1/2 such condition guarantees also the reduction property of the density matrices $\hat{\rho}^{i\in \nu} =\mathrm{Tr}_{j}\,\left(\hat{\rho}^{i\in \nu,j\in  \mu}\right)$ \cite{Brown1}. Thus, from Eq.(\ref{eq8}) we obtain the set of four variational equations:

\begin{eqnarray}
\tanh\left[\frac{\beta}{2}\left(\Lambda^{A}+h\right)\right]&=&\frac{\exp\left(\frac{\beta J^{AA}_{z}}{4}\right)\sinh\left[\beta\left(\Lambda^{AA}+h\right)\right]}{\exp\left(\frac{\beta J^{AA}_{z}}{4}\right)\cosh\left[\beta\left(\Lambda^{AA}+h\right)\right]+\exp\left(-\frac{\beta J^{AA}_{z}}{4}\right)\cosh\left(\frac{\beta J^{AA}_{\perp}}{2}\right)}
\label{eq9a}
\\
\tanh\left[\frac{\beta}{2}\left(\Lambda^{A}+h\right)\right]&=&\frac{\exp\left(\frac{\beta J^{AB}_{z}}{4}\right)\sinh\left[\beta\left(\Lambda^{AB}+h\right)\right]+\frac{\Delta^{AB}}{\sqrt{\left(\Delta^{AB}\right)^2+\left(J_{\perp}^{AB}\right)^2}} \exp\left(-\frac{\beta J^{AB}_{z}}{4}\right)\sinh\left(\frac{\beta\sqrt{\left(\Delta^{AB}\right)^2+\left(J_{\perp}^{AB}\right)^2}}{2}\right)}{\exp\left(\frac{\beta J^{AB}_{z}}{4}\right)\cosh\left[\beta\left(\Lambda^{AB}+h\right)\right]+\exp\left(-\frac{\beta J^{AB}_{z}}{4}\right)\cosh\left(\frac{\beta\sqrt{\left(\Delta^{AB}\right)^2+\left(J_{\perp}^{AB}\right)^2}}{2}\right)}
\label{eq9b}
\\
\tanh\left[\frac{\beta}{2}\left(\Lambda^{B}+h\right)\right]&=&\frac{\exp\left(\frac{\beta J^{BB}_{z}}{4}\right)\sinh\left[\beta\left(\Lambda^{BB}+h\right)\right]}{\exp\left(\frac{\beta J^{BB}_{z}}{4}\right)\cosh\left[\beta\left(\Lambda^{BB}+h\right)\right]+\exp\left(-\frac{\beta J^{BB}_{z}}{4}\right)\cosh\left(\frac{\beta J^{BB}_{\perp}}{2}\right)}
\label{eq9c}
\\
\tanh\left[\frac{\beta}{2}\left(\Lambda^{B}+h\right)\right]&=&\frac{\exp\left(\frac{\beta J^{AB}_{z}}{4}\right)\sinh\left[\beta\left(\Lambda^{AB}+h\right)\right]-\frac{\Delta^{AB}}{\sqrt{\left(\Delta^{AB}\right)^2+\left(J_{\perp}^{AB}\right)^2}} \exp\left(-\frac{\beta J^{AB}_{z}}{4}\right)\sinh\left(\frac{\beta\sqrt{\left(\Delta^{AB}\right)^2+\left(J_{\perp}^{AB}\right)^2}}{2}\right)}{\exp\left(\frac{\beta J^{AB}_{z}}{4}\right)\cosh\left[\beta\left(\Lambda^{AB}+h\right)\right]+\exp\left(-\frac{\beta J^{AB}_{z}}{4}\right)\cosh\left(\frac{\beta\sqrt{\left(\Delta^{AB}\right)^2+\left(J_{\perp}^{AB}\right)^2}}{2}\right)}
\label{eq9d}.
\end{eqnarray}

The set of Eqs.(\ref{eq9a}) - (\ref{eq9d}) has the same form both for the bilayer and multilayer system; however, the difference lies in the molecular field parameters $\Lambda$ and $\Delta$ which are different for those two cases (and are given by Eq.(\ref{eq6}) or Eq.(\ref{eq7}), respectively).
From the above variational equations the parameters $\lambda^{AA}, \, \lambda^{BB}, \, \lambda^{AB}$ and $\lambda^{BA}$ can be found numerically.

\section{Determination of the critical temperature}

One of the straightforward applications of variational equations (\ref{eq9a}) - (\ref{eq9d}) is the critical temperature calculation. For the continuous phase transitions the molecular field parameters should vanish. As a result of linearization of Eqs.(\ref{eq9a}) - (\ref{eq9d}) (when $\lambda^{\nu \mu} \to 0$) the critical temperature $T_c$ can be determined from the condition:
\begin{equation}
\det (\hat{M})=0
\label{eq10}
\end{equation}
where $\hat{M}$ is the characteristic matrix ($4 \times 4$) whose elements for the bilayer system can be written as:
\begin{eqnarray}
M_{11}&=&M_{34}=\beta_{c}\frac{\exp\left(\beta_{c}J^{AA}_{z}/2\right)-2\cosh\left(\beta_{c}J^{AA}_{\perp}/2\right)}{\exp\left(\beta_{c}J^{AA}_{z}/2\right)+\cosh\left(\beta_{c}J^{AA}_{\perp}/2\right)}\nonumber\\
M_{12}&=&M_{33}=\frac{\beta_{c}}{2}\frac{\exp\left(\beta_{c}J^{AA}_{z}/2\right)-\cosh\left(\beta_{c}J^{AA}_{\perp}/2\right)}{\exp\left(\beta_{c}J^{AA}_{z}/2\right)+\cosh\left(\beta_{c}J^{AA}_{\perp}/2\right)}\nonumber\\
M_{13}&=&M_{14}=M_{23}=M_{31}=M_{32}=M_{42}=0\nonumber\\
M_{21}&=&M_{44}=\frac{\left(4/J^{AB}_{\perp}\right) \sinh\left(\beta_{c}J^{AB}_{\perp}/2\right)-2\beta_{c} \cosh\left(\beta_{c}J^{AB}_{\perp}/2\right)}{\exp\left(\beta_{c}J^{AB}_{z}/2\right)+\cosh\left(\beta_{c}J^{AB}_{\perp}/2\right)}\nonumber\\
M_{22}&=&M_{43}=-\frac{\beta_{c}}{2}\nonumber\\
M_{24}&=&M_{41}=\frac{2\beta_{c} \exp\left(\beta_{c}J^{AB}_{\perp}/2\right)-\left(4/J^{AB}_{\perp}\right) \sinh\left(\beta_{c}J^{AB}_{\perp}/2\right)}{\exp\left(\beta_{c}J^{AB}_{z}/2\right)+\cosh\left(\beta_{c}J^{AB}_{\perp}/2\right)}\nonumber\\\
\label{eq11}
\end{eqnarray}
where $\beta_{c}=1/k_{\rm B}T_{c}$. 

For the multilayer system the matrix elements take the form of:
\begin{eqnarray}
M_{11}&=&\beta_{c}\frac{\exp\left(\beta_{c}J^{AA}_{z}/2\right)-2\cosh\left(\beta_{c}J^{AA}_{\perp}/2\right)}{\exp\left(\beta_{c}J^{AA}_{z}/2\right)+\cosh\left(\beta_{c}J^{AA}_{\perp}/2\right)}\nonumber\\
M_{12}&=&\beta_{c}\frac{\exp\left(\beta_{c}J^{AA}_{z}/2\right)-\cosh\left(\beta_{c}J^{AA}_{\perp}/2\right)}{\exp\left(\beta_{c}J^{AA}_{z}/2\right)+\cosh\left(\beta_{c}J^{AA}_{\perp}/2\right)}\nonumber\\
M_{13}&=&M_{14}=M_{31}=M_{32}=0\nonumber\\
M_{21}&=&M_{44}=\frac{\left(4/J^{AB}_{\perp}\right) \sinh\left(\beta_{c}J^{AB}_{\perp}/2\right)-2\beta_{c} \cosh\left(\beta_{c}J^{AB}_{\perp}/2\right)}{\exp\left(\beta_{c}J^{AB}_{z}/2\right)+\cosh\left(\beta_{c}J^{AB}_{\perp}/2\right)}\nonumber\\
M_{22}&=&M_{43}=\frac{\left(1/J^{AB}_{\perp}\right) \sinh\left(\beta_{c}J^{AB}_{\perp}/2\right)}{\exp\left(\beta_{c}J^{AB}_{z}/2\right)+\cosh\left(\beta_{c}J^{AB}_{\perp}/2\right)}-\frac{\beta_{c} \cosh\left(\beta_{c}J^{AB}_{\perp}/2\right)+\left(\beta_{c}/2\right)\exp\left(\beta_{c} J^{AB}_{z}/2\right)}{\exp\left(\beta_{c}J^{AB}_{z}/2\right)+\cosh\left(\beta_{c}J^{AB}_{\perp}/2\right)}\nonumber\\
M_{23}&=&M_{42}=\frac{\left(\beta_{c}/2\right)\exp\left(\beta_{c} J^{AB}_{z}/2\right)-\left(1/J^{AB}_{\perp}\right) \sinh\left(\beta_{c}J^{AB}_{\perp}/2\right)}{\exp\left(\beta_{c}J^{AB}_{z}/2\right)+\cosh\left(\beta_{c}J^{AB}_{\perp}/2\right)}\nonumber\\
M_{24}&=&M_{41}=\frac{2\beta_{c}\exp\left(\beta_{c} J^{AB}_{z}/2\right)-\left(4/J^{AB}_{\perp}\right) \sinh\left(\beta_{c}J^{AB}_{\perp}/2\right)}{\exp\left(\beta_{c}J^{AB}_{z}/2\right)+\cosh\left(\beta_{c}J^{AB}_{\perp}/2\right)}\nonumber\\
M_{33}&=&\beta_{c}\frac{\exp\left(\beta_{c}J^{BB}_{z}/2\right)-\cosh\left(\beta_{c}J^{BB}_{\perp}/2\right)}{\exp\left(\beta_{c}J^{BB}_{z}/2\right)+\cosh\left(\beta_{c}J^{BB}_{\perp}/2\right)}\nonumber\\
M_{34}&=&\beta_{c}\frac{\exp\left(\beta_{c}J^{BB}_{z}/2\right)-2\cosh\left(\beta_{c}J^{BB}_{\perp}/2\right)}{\exp\left(\beta_{c}J^{BB}_{z}/2\right)+\cosh\left(\beta_{c}J^{BB}_{\perp}/2\right)}
\label{eq12}
\end{eqnarray}

In a general case the determinant equation (\ref{eq10}) can be solved only numerically. In the particular case, when the layers $A$ and $B$ are magnetically equivalent, the determinant can be reduced to the form of eqs. (\ref{eq13}) or (\ref{eq14}). It should also be noted that the physical solution corresponding to the critical temperature of the system is the highest value of $T_c$ obtained from (\ref{eq10}).
\bibliographystyle{elsarticle-num}








\end{document}